%% file: youtu_template.tex
\documentclass[]{youtu} 

\usepackage{mathpazo}
\usepackage{graphicx}
\usepackage{natbib}
\usepackage{graphicx}
\usepackage{amssymb}
\usepackage{algorithm}
\usepackage{algorithmic}

\newcommand{\eg}{\textit{e.g.}}

\title{A Pipeline-Parallel System for Depth-Recurrent Models}
\author{Yunjia Xi\textsuperscript{1}, Jiarui Qin\textsuperscript{1}, Di Yin\textsuperscript{$1\dagger$}, Xing Sun\textsuperscript{1}}

\affiliation{$^1$Tencent Youtu Lab}

\hypersetup{
    pdftitle={A Pipeline-Parallel System for Depth-Recurrent Models},
    pdfauthor={Yunjia Xi, Jiarui Qin, Di Yin, Xing Sun},
    pdfsubject={Distributed, Parallel, and Cluster Computing},
    pdfkeywords={pipeline parallelism, depth-recurrent models, looped transformers,
                 mixture of experts, distributed training},
}

\begin{document}

\abstract{Depth-recurrent (Looped) models raise effective depth by applying a shared stack of blocks
several times. Scaling them usually requires pipeline parallelism, but commonly used
interleaved 1F1B is drawn for an acyclic stage graph in which each micro-batch enters
each stage once. Once that stack is split across stages, the same stage is visited 
several times on one micro-batch and later stages feed earlier ones: native 1F1B has 
no slot for the extra visit, cannot rank a first iteration against a second, 
and its neighbour-only send/receive order has no back edge to the first stage, 
so the arrangement stalls or fires out of order. We present Loop-PP, which rebuilds both 
the timetable and the communication for this cyclic layout. 
It schedules different iterations of the same stage separately, 
specializes a 1F1B skeleton into a globally consistent order over those visits, 
and emits a static instruction stream per rank that separates posting from waiting 
so communication overlaps local compute. Loop-PP is bit-identical to a single-stage reference. 
On 8B-60B MoE models, it lowers peak memory by as much as 32.8\%, and with compute budget fixed, 
the best observed throughput rises by 12.7\%.}
\maketitle

\begingroup
\renewcommand{\thefootnote}{}
\footnotetext{$\dagger$ Corresponding author. }
\endgroup

\vspace{-.1em}


\input{content/intro}
\input{content/method}
\input{content/exp}
\input{content/RW}
\section{Conclusion}
We identified a fundamental mismatch between conventional PP and looped Transformers: 
repeated cross-stage traversal introduces recurrent dependencies that independent stage-local repetition cannot preserve. 
We presented \emph{Loop-PP}, to our knowledge the first pipeline-parallel training system for looped Transformers. 
Its loop-aware compiler represents logical recurrent visits independently of physical parameter storage and compiles them into dependency-aware Global Compute Orders and static Per-Rank Action Streams. 
Across 8B--60B models, Loop-PP reduces peak memory by up to 32.8\% 
and improves the best measured throughput among the evaluated configurations by up to 12.7\% under an equal resource budget. 
Our analysis further shows that its performance depends on recurrent placement and compute-chunk granularity, 
favoring layouts with sufficient independent local work.

\setcitestyle{numbers,square}
\setcitestyle{square,numbers,comma}
{\footnotesize
\setlength{\bibsep}{2pt plus 0.3pt}
\bibliography{youtu_bib}
}

\appendix
\input{content/appendix}
\end{document}

%% file: content/intro.tex
\section{Introduction}
\label{sec:introduction}
Recent advances in large language models (LLMs) have been driven primarily by scaling model parameters~\citep{kaplan2020scaling,brown2020gpt3}, data~\citep{hoffmann2022chinchilla}, and test-time computation~\citep{snell2024testtime}.
Looped Transformers or depth-recurrent Transformers offer a different path by recursively applying the same blocks, increasing effective model depth and capacity without adding parameters~\citep{dehghani2019universal,lan2020albert,geiping2025huginn,zhu2025ouro}.
Recent studies have shown that such models hold promise for reasoning, in-context learning, and test-time scaling~\citep{saunshi2025reasoning,yang2024looped,alabdulmohsin2025rins}.

Existing work, however, has focused primarily on model architectures, leaving the systems support for large-scale depth-recurrent training underexplored.
As models outgrow single-device memory, their parameters must be distributed across devices. 
Pipeline parallelism (PP) partitions model layers into stages placed across devices and pipelines micro-batches through them, 
offering favorable communication volume and arithmetic intensity for large-scale training~\citep{huang2019gpipe,narayanan2021pipedream2bw,shoeybi2019megatron,rajbhandari2020zero}.

\begin{figure}[t]
    \centering
    \includegraphics[width=0.8\linewidth]{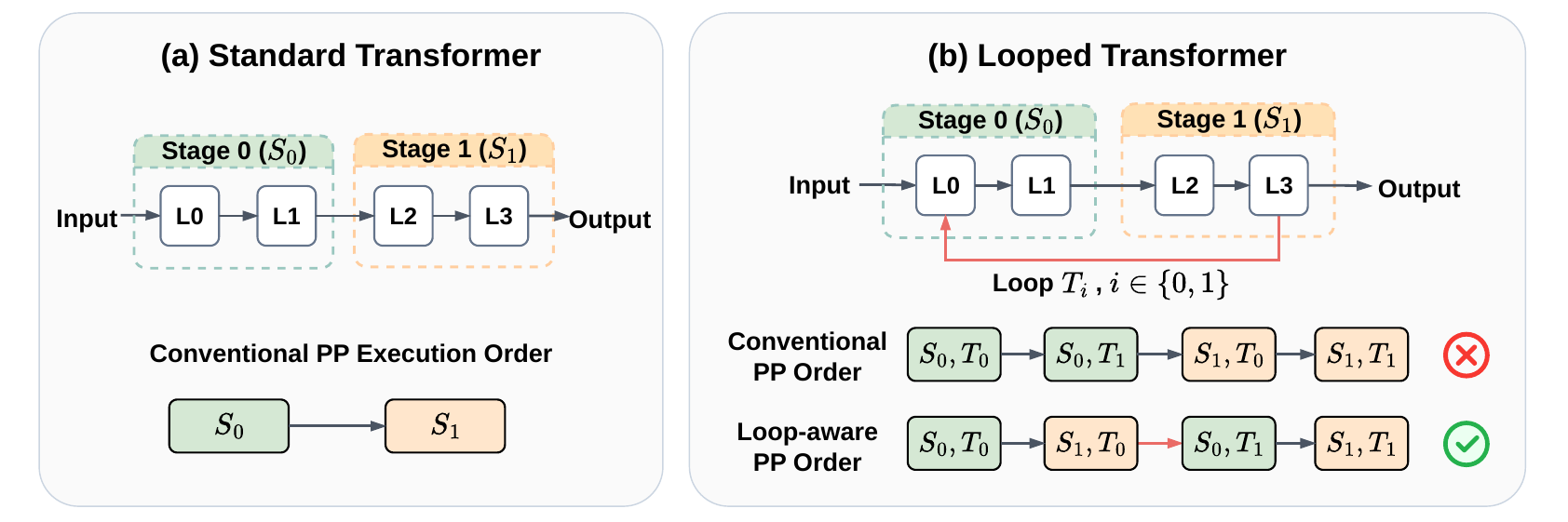}
    \caption{Pipeline execution of (a) Standard Transformer and (b) Looped Transformer.}
    \label{fig:motivation}
\end{figure}

However, conventional PP schedules, \eg, Megatron's interleaved 1F1B~\citep{narayanan2021efficient}, are designed for 
linear feed-forward execution, where each layer is used once in a forward pass.
Looped Transformers instead revisit shared blocks spanning multiple stages, 
introducing loop-carried feedback from later to earlier stages that breaks the compute-readiness 
and communication-order assumptions of native interleaved 1F1B (Figure~\ref{fig:motivation}). 
Without global rescheduling, the direct adaptation of conventional PP is limited to independent stage-local repetition.
Consider a recurrent region repeated for $\mathcal{T}$ steps and partitioned across $P$ stages, 
where $\mathcal{F}_p$ denotes the recurrent computation assigned to stage $p$. 
Such stage-local repetition yields
\begin{equation}
    \underbrace{
        \mathcal{F}_{P-1}^{\mathcal{T}}\circ\cdots\circ\mathcal{F}_0^{\mathcal{T}}
    }_{\text{independent stage-local repetition}},
    \label{eq:local-recurrence}
\end{equation}
whereas the looped model requires
\begin{equation}
    \underbrace{
        \left(\mathcal{F}_{P-1}\circ\cdots\circ\mathcal{F}_0\right)^{\mathcal{T}}
    }_{\text{$\mathcal{T}$ end-to-end recurrent steps}}.
    \label{eq:global-recurrence}
\end{equation}
Here, the superscript $\mathcal{T}$ denotes repeated function composition.
Thus, simply inserting an independent local loop into each pipeline stage is insufficient.

A straightforward alternative is to physically unroll the $\mathcal{T}$ 
recurrent visits into separate pipeline chunks and apply conventional PP scheduling. 
However, using independent parameters changes the model semantics, 
while tying the expanded copies incurs additional parameter storage and synchronization overhead. 
The challenge is therefore to preserve exact recurrence semantics 
and a single physical copy of shared parameters while retaining the concurrency 
of native pipeline execution.

To this end, we present \emph{Loop-PP}, the first pipeline-parallel training system for looped Transformers.
Its core is a \emph{Loop-Aware Pipeline Compiler} which translates recurrent model semantics into static pipeline execution plans while preserving a single physical copy of shared parameters.
First, \emph{Semantic Graph Construction} logically unrolls the recurrent computation into distinct visits to shared physical model regions and captures their cross-step dependencies, yielding a scheduling-independent \emph{Semantic Graph}.
Second, \emph{Native-Guided Schedule Synthesis} uses native interleaved 1F1B as a compute template and constrains it with these semantic dependencies, producing a \emph{Global Compute Order}.
Third, \emph{Communication-Aware Stream Compilation} turns cross-stage dependencies into ordered communication actions and schedules their initiation and completion around dependency-ready compute to expose communication overlap, producing a static \emph{Per-Rank Action Stream} for each rank.
We further introduce system optimizations to reduce communication and runtime overhead. In summary, our main contributions are:
\begin{itemize}
    \item We identify the mismatch between conventional pipeline parallelism and depth-recurrent execution, and, to our knowledge, present the first pipeline-parallel training system designed for looped Transformers.   
    \item We develop Loop-PP, a loop-aware pipeline compiler that preserves recurrent semantics and parameter sharing while adapting native interleaved 1F1B scheduling and communication to repeated cross-stage execution.
    \item Across 8B--60B models, Loop-PP maintains broadly comparable throughput to compute-matched baseline with interleaved 1F1B while reducing peak memory by up to 32.8\%. Under an equal resource budget, it improves the best measured throughput by up to 12.7\%.
\end{itemize}



%% file: content/method.tex
\section{Background and Motivation}
\label{sec:background}

\paragraph{Looped Transformers}
\label{sec:background-recurrent}

Looped models increase effective depth by repeatedly applying shared blocks~\citep{dehghani2019universal,geiping2025huginn,zhu2025ouro,saunshi2025reasoning}.
Existing models vary in recurrence scope, from intermediate regions~\citep{geiping2025huginn} to nearly the entire Transformer backbone~\citep{zhu2025ouro}.
To capture both cases, we decompose the model into a non-recurrent prelude $\mathcal{P}$, a recurrent region $\mathcal{R}$, and a non-recurrent coda $\mathcal{C}$.
For full-backbone recurrence, $\mathcal{P}$ and $\mathcal{C}$ contain only boundary computation such as the embedding and output head.

Let $a = \mathcal{P}(x;\theta_{\mathcal{P}})$ denote the prelude output with input $x$ and parameters $\theta_{\mathcal{P}}$.
Starting from $h_0=a$, the recurrent region is applied for $\mathcal{T}$ steps:
\begin{equation}
    h_{t+1}=
    \mathcal{R}(h_t,a;\theta_{\mathcal{R}}),
    \qquad
    t=0,\ldots,\mathcal{T}-1,
    \label{eq:depth-recurrent-model}
\end{equation}
followed by $y = \mathcal{C}(h_{\mathcal{T}};\theta_{\mathcal{C}})$.
The parameters $\theta_{\mathcal{R}}$ are shared across recurrent steps, 
and the dependence of $h_{t+1}$ on $h_t$ captures the recurrent state carried across steps.
The second argument $a$ is optional: some models omit it and update from $h_t$ alone, 
while others~\citep{geiping2025huginn} inject $a$ at every step, making the unrolled computation a DAG rather than a linear chain.
When these dependencies span pipeline stages, PP must communicate recurrent states and, 
when present, injected inputs in addition to the hidden activations transferred in conventional feed-forward execution.
Conventional PP assumes a single feed-forward visit to each model chunk, whereas a recurrent region spanning stages requires cross-stage feedback and side dependencies.
Therefore, we need to design a new pipeline parallel scheduling strategy for looped models.

\section{Loop-Aware Pipeline Compiler}
\label{sec:compiler}

\begin{figure*}[t]
    \centering
    \includegraphics[width=\textwidth]{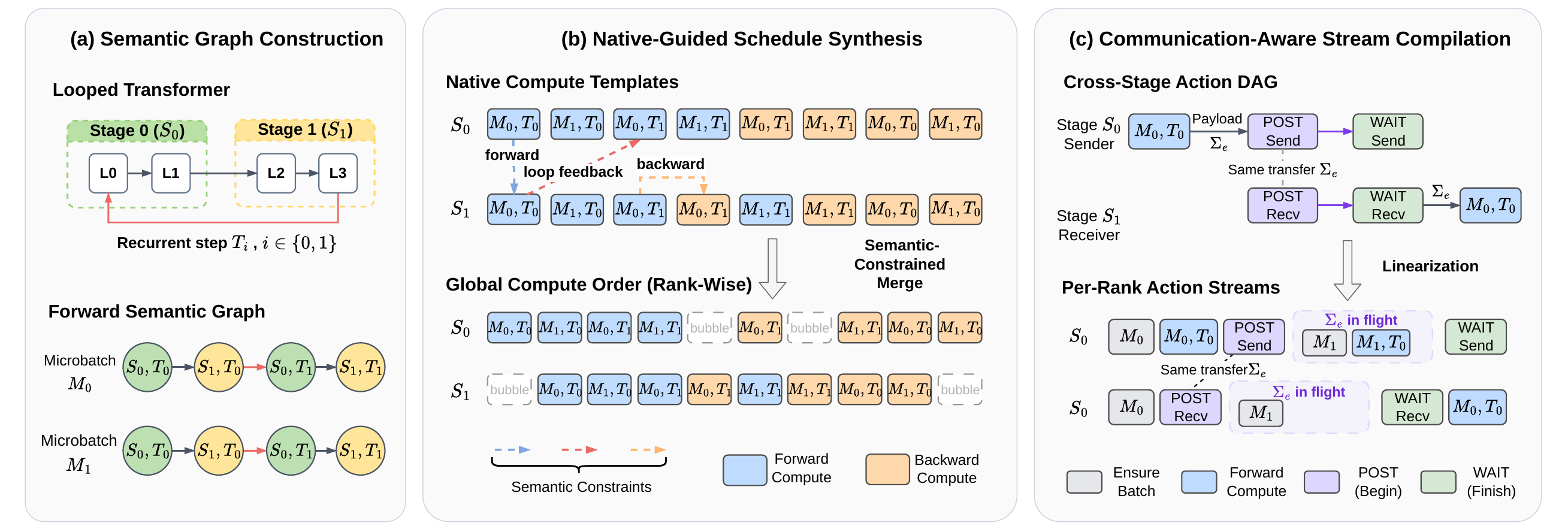}
    \caption{
    Overview of the Loop-Aware Pipeline Compiler. $S_p$ denotes pipeline stage $p$, $T_t$ denotes recurrent step $t$, and $M_m$ denotes micro-batch $m$.
    }
    \label{fig:compiler-overview}
\end{figure*}

The \emph{Loop-Aware Pipeline Compiler} translates a depth-recurrent model into deterministic computation and communication sequences for pipeline-parallel training.
It takes as input the recurrent model structure, the number of micro-batches $M$, and PP/VPP placement, and produces static Per-Rank Action Streams for each pipeline rank through the following three stages, as shown in Figure~\ref{fig:compiler-overview}.

\subsection{Semantic Graph Construction}
\label{sec:semantic-graph}

A looped model repeatedly invokes the same physical blocks, but each invocation consumes a different intermediate state and therefore has a distinct logical identity.
Figure~\ref{fig:compiler-overview} (a) makes this unrolling explicit: $T_0$ and $T_1$ on the same pipeline stage $S_p$ are two logical visits to the same physical module, while different micro-batches $M_m$ form disjoint graph copies.
The compiler unrolls the fixed $\mathcal{T}$ recurrent steps into a forward \emph{Semantic Graph} $\mathcal{G}=(\mathcal{V},\mathcal{E})$.
For each micro-batch, the dataflow has the form
\begin{equation}
    \mathcal{P}
    \rightarrow
    \mathcal{R}^{(0)}
    \rightarrow \cdots \rightarrow
    \mathcal{R}^{(\mathcal{T}-1)}
    \rightarrow
    \mathcal{C},
    \label{eq:semantic-unrolling}
\end{equation}
where each $\mathcal{R}^{(t)}$ is the within-step DAG of the recurrent region.

Concretely, the compiler first partitions the prelude, recurrent region, and coda into segments by PP/VPP placement and creates one semantic node $v$ for each visit.
Each node $v\in\mathcal{V}$ records its micro-batch $M_m$, recurrent step $T_t$, computation role, PP stage $S_p$, and VPP placement.
Prelude and coda are visited once per micro-batch, whereas the recurrent region contributes a new set of logical nodes at every $t$.
The compiler then connects computations within a step by main edges, connects the exit of step $t$ to the entry of step $t+1$ by feedback edges 
(the cross-stage recurrence in Figure~\ref{fig:compiler-overview}), and routes loop-invariant or auxiliary state $a$ to its consumers by injection edges.
An edge $e=(u,v,\Sigma_e)$ carries an ordered tensor bundle $\Sigma_e$ from producer $u$ to consumer $v$ that may contain multiple tensors, \eg, auxiliary state $a$ and main hidden activation.

Unrolling creates a distinct logical visit for each recurrent step, so a feedback edge connects step $t$ only to step $t+1$ and the graph is acyclic for finite $\mathcal{T}$.
Injection edges may additionally fan out from the prelude to several recurrent nodes, so $\mathcal{G}$ need not be a single chain.
The graph specifies what to compute, which dependencies to respect, and on which $S_p$ each node runs, but it does not yet choose an execution order.
Backward compute is derived later from the reversed dependencies.
Appendix~\ref{app:semantic-graph} gives the complete node and edge definitions and the semantic-equivalence argument.

\subsection{Native-Guided Schedule Synthesis}
\label{sec:global-compute-scheduling}

The Semantic Graph defines a partial order over model computations, but it does not specify which operation to choose when several are simultaneously ready.
Any topological execution preserves model semantics, yet an arbitrary choice can leave pipeline stages idle.
Efficient scheduling must satisfy the recurrent dependencies while retaining the compute cadence of established pipeline schedules~\citep{huang2019gpipe,narayanan2021pipedream2bw,qi2024zerobubble}.
We therefore use native interleaved 1F1B~\citep{narayanan2021efficient} as a compute template and constrain it with the Semantic Graph.
This section orders forward and backward computation without introducing communication.

\paragraph{Native Compute Template.}
For each PP stage, the compiler treats every visit to a layer-bearing segment on $S_p$ as a \emph{logical compute chunk}: recurrent visits are distinct chunks but still invoke the same physical parameters.
It then applies Megatron's interleaved 1F1B rule to build a communication-free reference sequence for each stage,
\begin{equation}
    \mathcal{N}^{\mathrm{nat}}_p
    =
    [(\phi_i,m_i,c_i)]_{i=0}^{J_p-1},
    \qquad \phi_i\in\{F,B\},
    \label{eq:native-compute-template}
\end{equation}
where $J_p$ is the number of native compute slots on stage $p$.
Slot $i$ records the phase $\phi_i$ (forward or backward), the micro-batch $m_i$ assigned to that slot, and the logical compute chunk $c_i$.
Figure~\ref{fig:compiler-overview} (b) illustrates that $T_0$ and $T_1$ on the same $S_p$ are two logical visits, while $M_m$ indexes the forward/backward slots of different micro-batches.
Recurrence therefore expands the number of logical slots, not the number of physical layers.

\paragraph{Semantic-constrained Merge.}
The native templates specify the desired interleaved 1F1B cadence, but do not encode recurrent dependencies in the Semantic Graph $\mathcal{G}$.
Executing them independently may therefore start a semantic consumer before its input has been produced.
To make these dependencies explicit, the compiler expands each native slot into the semantic nodes attached to its logical compute chunk and 
represents every node $v$ by two phased semantic operations, forward operation $(F,v)$ and backward operation $(B,v)$.
These semantic operations, rather than the abstract native slots, are the objects ultimately ordered by the scheduler.

The compiler then combines the semantic dependencies in $\mathcal{G}$ with the native order $\mathcal{N}^{\mathrm{nat}}_p$ into a phased constraint graph,
\begin{equation}
    \mathcal{H}
    =
    \left(
        \{(\phi,v)\mid \phi\in\{F,B\},\,v\in\mathcal{V}\},
        \mathcal{E}_{\mathrm{nat}}
        \cup \mathcal{E}_{F}
        \cup \mathcal{E}_{B}
        \cup \mathcal{E}_{FB}
    \right),
    \label{eq:phased-constraint-graph}
\end{equation}
For every semantic edge $(u,v)\in\mathcal{E}$, $\mathcal{E}_{F}$ contains $(F,u)\prec(F,v)$ and $\mathcal{E}_{B}$ contains the reverse-mode constraint $(B,v)\prec(B,u)$.
For every node $v\in\mathcal{V}$, $\mathcal{E}_{FB}$ contains $(F,v)\prec(B,v)$.
The Semantic Constraints highlighted in Figure~\ref{fig:compiler-overview} illustrate these model-derived dependencies.
In particular, the illustrated loop feedback is an edge in $\mathcal{E}_{F}$ requiring the exit forward of step $t$ to precede the entry forward of step $t+1$.
On each stage, native 1F1B requires operations to follow their order in $\mathcal{N}^{\mathrm{nat}}_p$.
After mapping its slots to phased semantic operations, the compiler encodes this order by adding an edge between each consecutive pair,
and their union across all stages forms $\mathcal{E}_{\mathrm{nat}}$.

Next, the compiler computes a deterministic linear extension of $\mathcal{H}$ with a priority-guided Kahn algorithm~\citep{kahn1962topological}.
All zero-indegree operations form a global ready heap, so an operation becomes eligible only after all of its semantic and stage-local native predecessors have been emitted.
Among them, the compiler selects the lexicographically smallest $(r_{\mathrm{nat}},r_{\mathrm{phase}},\operatorname{id})$, where $r_{\mathrm{nat}}$ is its position in the native order of its stage, $r_{\mathrm{phase}}=0$ for forward and $1$ for backward, and the semantic-node identity provides deterministic tie breaking.
Because $\mathcal{E}_{\mathrm{nat}}$ chains the operations on each stage, the algorithm chooses only among the ready frontiers of different stages without changing any stage-local native order.
Emitting all operations produces the \emph{Global Compute Order}
\begin{equation}
    \mathcal{O}
    =
    [(\phi_0,v_0),\ldots,(\phi_{N_{\mathrm{op}}-1},v_{N_{\mathrm{op}}-1})],
    \qquad \phi_i\in\{F,B\},\quad v_i\in\mathcal{V},\quad
    N_{\mathrm{op}}=2|\mathcal{V}|,
    \label{eq:global-phased-compute-order}
\end{equation}
which contains every semantic forward and backward operation exactly once. 
As illustrated by the rank-wise view in Figure~\ref{fig:compiler-overview}(b), the Global Compute Order preserves 
each rank's native compute order without serializing independent operations across ranks at runtime.
Native-rank priority is applied only among zero-indegree operations in $\mathcal{H}$, so it never bypasses a semantic dependency.

Under balanced chunk costs, the structural bubble matches that of compute-matched interleaved 1F1B.
Note that structural bubbles characterize only the compute schedule.
The actual execution time also depends on computation-communication overlap, which we address next.

\subsection{Communication-Aware Stream Compilation}
\label{sec:per-rank-compilation}

The Global Compute Order specifies the computation order, but it does not determine communication actions.
This section introduces communication dependencies and compiles the compute and communication actions into a deterministic executable stream for every pipeline rank.

\paragraph{Communication Planning.}
The compiler classifies each semantic edge by the PP placements of its endpoints: a local edge whose endpoints share a stage remains an in-memory data dependency, 
whereas a remote edge spanning stages induces a planned transfer.
For a remote edge $e=(u,v,\Sigma_e)$, a forward transfer moves $\Sigma_e$ from the stage owning $u$ to the stage owning $v$, whereas a backward transfer moves $\nabla\Sigma_e$ from the stage owning $v$ to the stage owning $u$. 
A transfer is represented as
\begin{equation}
    \gamma
    =
    (e,\phi,s_{\mathrm{src}},s_{\mathrm{dst}},\Sigma_e^\phi),
    \qquad \phi\in\{F,B\},
    \label{eq:planned-transfer}
\end{equation}
where $\phi$ denotes the forward or backward direction, $s_{\mathrm{src}}$ and $s_{\mathrm{dst}}$ identify the communicating stages, 
and $\Sigma_e^F=\Sigma_e$ and $\Sigma_e^B=\nabla\Sigma_e$ denote the corresponding payload schemas.
Remote transfers are ordered according to the forward and backward subsequences of $\mathcal{O}$.
The resulting \emph{Communication Plan} specifies what data moves between which stages and in what order.

\paragraph{Cross-Stage Action DAG.}
To represent communication--computation overlap, the compiler separates the initiation and completion of every transfer.
A \textsc{POST} submits the nonblocking send or receive bundle and returns before the transfer necessarily completes, 
whereas the corresponding \textsc{WAIT} confirms the transfer's completion.
Thus, it expands $\mathcal{O}$ and the Communication Plan into a \emph{Cross-Stage Action DAG} containing forward/backward \textsc{Compute}, nonblocking \textsc{POST}, \textsc{WAIT}, and \textsc{EnsureBatch} for micro-batch preparation.

For a transfer $\gamma$ associated with $e=(u,v,\Sigma_e)$, let $(x,y)=(u,v)$ when $\phi=F$ and $(x,y)=(v,u)$ when $\phi=B$.
The Action DAG enforces
\begin{equation}
    \begin{aligned}
        \operatorname{Compute}^{\phi}(x)
        &\prec \operatorname{POST}_{\mathrm{src}}(\gamma),\\
        \left\{
            \operatorname{POST}_{\mathrm{src}}(\gamma),
            \operatorname{POST}_{\mathrm{dst}}(\gamma)
        \right\}
        &\prec
        \left\{
            \operatorname{WAIT}_{\mathrm{src}}(\gamma),
            \operatorname{WAIT}_{\mathrm{dst}}(\gamma)
        \right\},\\
        \operatorname{WAIT}_{\mathrm{dst}}(\gamma)
        &\prec \operatorname{Compute}^{\phi}(y).
    \end{aligned}
    \label{eq:remote-edge-actions}
\end{equation}
These constraints ensure that a payload is produced before transfer initiation, both endpoints initiate matching communication before completion, 
and the receiver completes the transfer before consuming the payload.
The graph also retains semantic compute dependencies and communication-order and resource-safety constraints required for execution.
Crucially, the Cross-Stage Action DAG specifies which action reorderings and overlaps are legal.
Actions with no dependency path between them may change relative order during linearization. 
In particular, independent computation may be placed between \textsc{POST} and \textsc{WAIT} to overlap with in-flight communication.
Figure~\ref{fig:compiler-overview}(c) illustrates how a remote dependency expands into POST and WAIT actions at its sender and receiver.

\paragraph{Native-Aware Communication Linearization.}
Because the Cross-Stage Action DAG defines only a partial order, it generally admits multiple legal executions.
Linearization selects deterministic Per-Rank Action Streams from these executions while retaining the native interleaved 1F1B compute cadence and exposing communication overlap.

Ranks cannot make this choice independently: selecting sends and receives in different orders may produce inconsistent communication matching.
The compiler therefore first fixes a global matching order for remote transfers.
Among ready transfers, it preferentially groups opposite-direction transfers between the same stages into a bidirectional communication \emph{epoch}, 
following the P2P structure of interleaved pipeline execution~\citep{narayanan2021efficient}. A transfer without a legal partner forms a singleton epoch.
An epoch represents one globally agreed communication-matching decision, not a runtime barrier.
Once their order is fixed, both endpoints initiate matching operations consistently without having to reach an epoch simultaneously.

The compiler then projects the fixed epoch order onto each stage and linearizes its local actions using three rules:
(i) post the next dependency-ready and resource-feasible epoch as early as possible;
(ii) otherwise, select the earliest ready \textsc{Compute} in that stage's native compute order; and
(iii) emit a \textsc{WAIT} only when a consumer requires the result or an in-flight transfer blocks further progress.
Consequently, POSTs open communication windows early, WAITs close them only when necessary, and independent native-preferred computation fills the interval between them.
The Action DAG guarantees legality, the epoch order guarantees consistent cross-rank matching, and these rules determine the concrete local order.
As illustrated in the lower part of Figure~\ref{fig:compiler-overview}(c), the result is
\begin{equation}
    \{\mathcal{Q}_s\}_{s=0}^{P-1},
    \label{eq:per-rank-action-streams}
\end{equation}
where $\mathcal{Q}_s$ is the \emph{Per-Rank Action Stream} for stage $s$.
All streams are compiled before execution and the runtime follows them directly.
Appendix~\ref{app:per-rank-compilation} proves that the compiled streams preserve the Semantic Graph dependencies and yield PP-transport deadlock freedom under explicit backend-progress assumptions.

\paragraph{Implementation Optimizations}
The three compilation stages above produce a correct and executable plan, but
steady-state training can still incur avoidable communication and compilation
overheads.
First, a uniform payload may carry loop auxiliary tensors even when remote consumers do not require
those fields.
\emph{Payload Localization} therefore rewrites each payload schema according to its
actual consumers and transmits only the fields required by remote computation.
Second, the Global Compute Order, Communication Plan, and Per-Rank Action Streams can be reused across training steps.
The compiler therefore uses \emph{compilation caching} to reuse these results, keyed by the Semantic Graph and execution configuration.
On a cache hit, steady-state training directly executes the validated local streams.

%% file: content/exp.tex
\section{Experiments}
\label{sec:experiments}

\subsection{Experimental Setup}
\label{sec:experimental-setup}

\paragraph{Models.}
We evaluate three full-loop MoE models based on Megatron-LM~\citep{narayanan2021efficient}:
8B-A1B, 30B-A3B, and 60B-A5B with 32, 48, and 48 unique layers.
For looped models, the complete layer stack is recurrently executed for $\mathcal{T}=2$ steps, giving effective depths of 64, 96, and 96 layers.
The model names refer to the unique physical parameterization rather than the parameter count of an untied expansion.
Unless otherwise specified, experiments use 32 GPUs, sequence length 4,096, and $\mathcal{T}=2$.
We report end-to-end training throughput in tokens per second and peak GPU memory in GiB.
On an 8B MoE model with $\mathcal{T}=2$, Loop-PP at PP2 matches a PP1 recurrent oracle bit-exactly in language-model and auxiliary loss over five steps.
After 1,000 steps with $\mathcal{T}=3$, the losses differ by $5.43\times10^{-4}$ (1.813975 vs.\ 1.814518), with no accumulating drift.

\paragraph{Baselines.}
We compare three strategies with Loop-PP.
\textit{Unrolled} materializes the same effective depth with independent parameters and schedules it with Megatron's interleaved 1F1B. 
It matches the computation of the looped models but not their parameter-sharing semantics, 
serving as an optimistic native-schedule reference.
\textit{Tied} uses the same physical expansion but ties the weights across recurrent visits, 
preserving the looped model's semantics. 
After backward, gradients are reduced, only one replica is optimized, 
and the updated weights are copied to all replicas. 
This preserves interleaved 1F1B scheduling at the cost of replicated parameters and additional synchronization.
\textit{Lockstep} executes the same Semantic Graph as Loop-PP with a conservative, 
dependency-respecting 1F1B schedule that bounds the number of live micro-batches. 
It serves as a correctness-oriented reference without Loop-PP's optimized scheduling and communication.

\subsection{Overall Performance}
\label{sec:overall-performance}

\begin{figure*}[htbp]
    \centering
    \includegraphics[width=\textwidth]{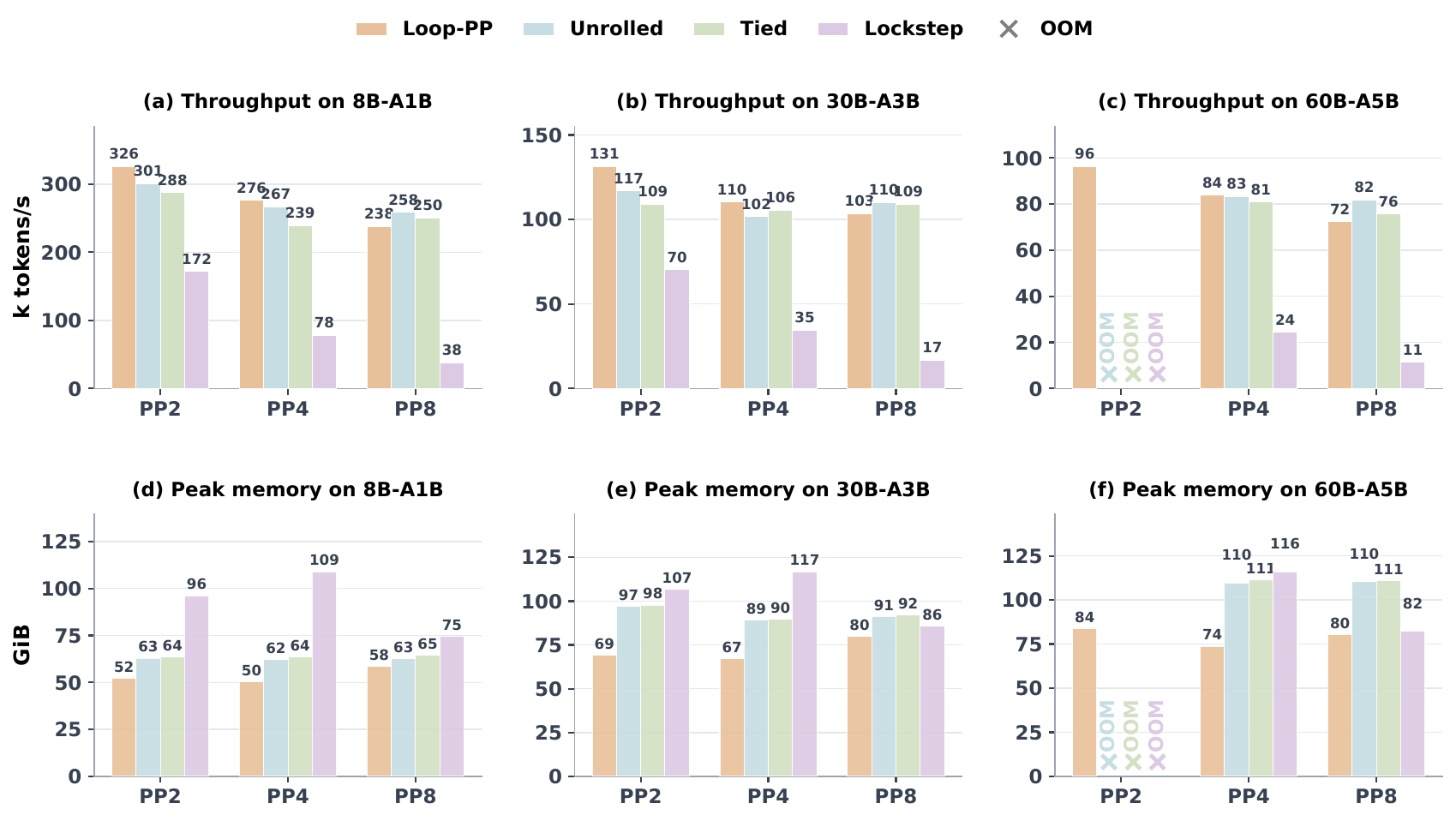}
    \caption{
    Training throughput and peak memory for three MoE models at PP2, PP4, and PP8.
    }
    \label{fig:end-to-end-performance}
\end{figure*}

\paragraph{Comparison under matched configuration.}
We evaluate Loop-PP and the baselines on different model sizes and PP degrees, 
and report throughput and peak memory in Figure~\ref{fig:end-to-end-performance}.
For a given model and PP degree, all methods use the same training configuration. 
The settings may slightly differ across models and PP degrees.

Loop-PP uses less peak memory than Unrolled and Tied at every plotted point, saving up to 32.8\% memory.
At 60B PP2, Unrolled, Tied, and Lockstep run out of memory, while Loop-PP completes.
Loop-PP's throughput is competitive with the expanded native schedules and depends on pipeline granularity.
At PP2, Loop-PP is 12.4\% faster than Unrolled on 30B, but at PP8 it is slower.
We will further analyze this reversal in Section~\ref{sec:in-depth-analysis}.
Tied is close to Unrolled but usually slightly slower, due to the extra gradient/parameter synchronization.
Relative to Lockstep, Loop-PP is $1.9\times$ faster at PP2 and more than $6\times$ faster at PP8 on 30B.
Thus Loop-PP approaches Unrolled in throughput, remains substantially faster than the conservative Lockstep baseline, 
and retains a large memory advantage.

\paragraph{Loop-PP achieves best performance under an equal resource budget.}


Unlike the previous comparison with the configuration fixed for each 
model and pipeline degree, we keep the model (Dense 14B or 30B MoE), the global batch size (512), 
and the 32-GPU budget fixed, and let Loop-PP and Unrolled choose different configurations, \eg, PP, VPP, micro-batch size, 
to saturate the 32 GPUs.
Table~\ref{tab:best-equal-resource} reports the best measured throughput of each method. 
On 30B-A3B, Loop-PP improves the best measured throughput by 12.7\%, 
using only slightly more peak memory at that throughput-optimal point. 
This shows that Loop-PP's memory saving is usable in practice: the saved memory 
can be converted into a more efficient operating point, so Loop-PP trains faster than 
the compute-matched Unrolled baseline under the same budget.

\begin{table}[htbp]
\centering
\small
\caption{Best throughput under an equal 32-GPU resource budget. Memory is peak allocated GiB.}
\label{tab:best-equal-resource}
\begin{tabular}{lrrrrr}
\hline
Model & Loop-PP & Unrolled & Speedup & Loop mem. & Unrolled mem. \\
\hline
Dense 14B & 96.6k & 87.0k & 1.110$\times$ & 113.7 & 115.7 \\
MoE 30B   & 170.7k & 151.5k & 1.127$\times$ & 117.7 & 112.7 \\
\hline
\end{tabular}
\end{table}

\subsection{In-Depth Analysis}
\label{sec:in-depth-analysis}
\begin{figure*}[htbp]
    \centering
    \includegraphics[width=\textwidth]{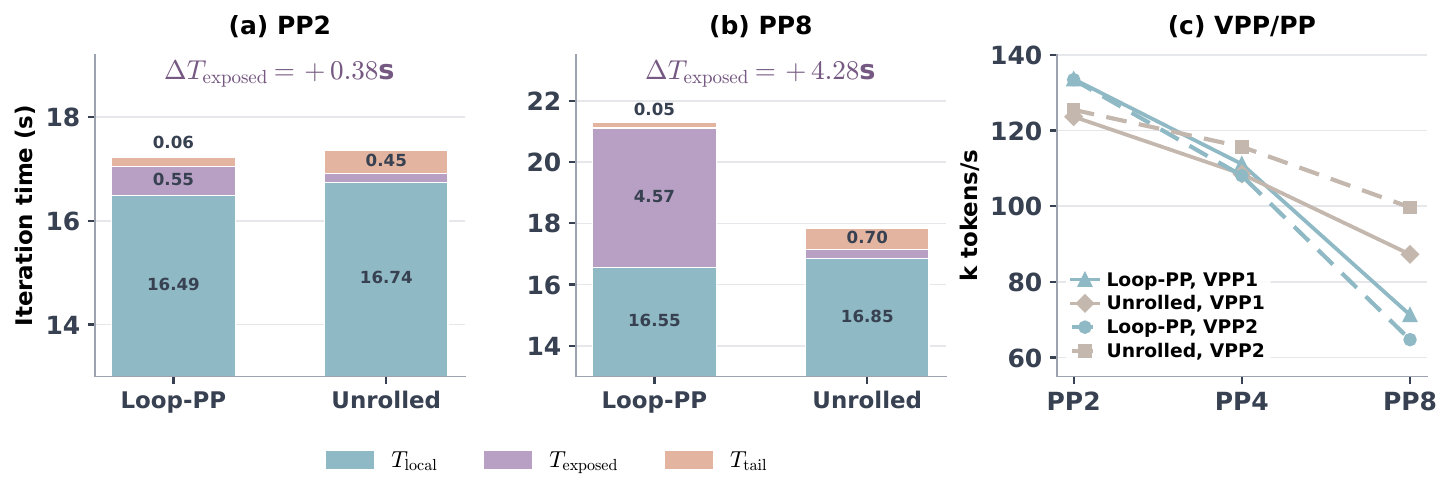}
    \caption{ Diagnosing the PP8 performance cliff from 30B MoE profiling.
    (a--b) Iteration decomposition at PP2 and PP8.
    (c) Throughput under different VPP/PP.
    }
    \label{fig:pp8-cliff}
\end{figure*}

\paragraph{Understanding the PP8 Performance Cliff}
To understand why Loop-PP falls behind at PP8, we profile matched 30B Loop-PP and Unrolled runs on 16 GPUs. 
We decompose iteration time $T_{\mathrm{step}}$ into three components. 
$T_{\mathrm{local}}$ is the service time of local forward and backward actions, including computation and work internal to MoE actions. 
$T_{\mathrm{exposed}}$ is pipeline time not covered by local actions and can be viewed approximately as the exposed pipeline cost.
$T_{\mathrm{tail}}$ contains gradient finalization, optimizer work, and other costs outside the pipeline schedule. 
Figure~\ref{fig:pp8-cliff}(a--b) presents this decomposition on PP2 and PP8. 
We additionally compare Loop-PP and Unrolled across VPP1 and VPP2 to study the effect of compute-chunk granularity in Figure~\ref{fig:pp8-cliff}(c).

In Figure~\ref{fig:pp8-cliff}(a--b), Loop-PP trades lower local and tail costs for a higher exposed pipeline cost. 
Its smaller $T_{\mathrm{local}}$ and $T_{\mathrm{tail}}$ are consistent with reusing shared parameters, 
while recurrent cross-stage dependencies increase $T_{\mathrm{exposed}}$, the exposed recurrent pipeline cost.
At PP2, Loop-PP has a thick compute chunk, so its  
shorter local computation and tail outweigh the additional exposed time.
At PP8, thinner compute chunks expose recurrent dependencies more frequently, 
increasing communication intensity and leaving less independent computation to hide it. 
Consequently, the exposed-time penalty overwhelms the local and tail savings, 
making Loop-PP slower than Unrolled.
Figure~\ref{fig:pp8-cliff}(c) further confirms this, 
where changing from VPP2 to VPP1 doubles the Loop-PP chunk from three to six unique layers
and raises its throughput from 64.7k to 71.3k tokens/s, narrowing the gap to Unrolled.
Together with the observation that 99.98\% of the measured PP8 receive-wait time is dependency-limited, 
this result shows that sufficiently thick compute chunks are important for amortizing 
the recurrent pipeline overhead, and Loop-PP should avoid overly fine pipeline partitions.

\begin{figure*}[htbp]
    \centering
    \includegraphics[width=0.86\textwidth]{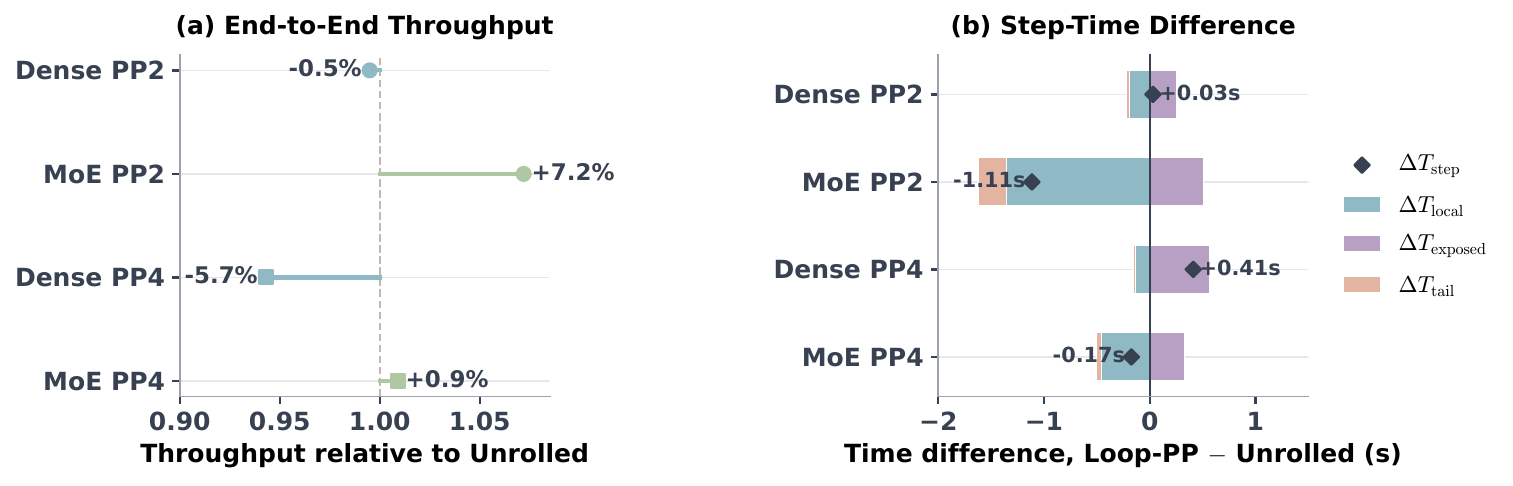}
    \caption{
    Comparison of Dense and MoE models.
    (a) Throughput relative to Unrolled.
    (b) Step-time difference $\Delta T_{\mathrm{step}}=T_{\mathrm{step}}^{\mathrm{Loop\text{-}PP}}-T_{\mathrm{step}}^{\mathrm{Unrolled}}$. The diamonds denote the total difference.
    }
    \label{fig:dense-moe-breakdown}
\end{figure*}

\paragraph{Loop-PP across Dense and MoE Workloads.}
To study workload sensitivity, we replace the MoE blocks in the 30B model with dense FFNs while otherwise retaining its architecture and activation settings, 
and evaluate both variants at PP2 and PP4 on 32 GPUs.
As shown in Figure~\ref{fig:dense-moe-breakdown}, Loop-PP reduces local-action and tail times 
but increases exposed pipeline time across all four settings.
The Dense results show that the local savings are not solely due to MoE load balancing and may arise 
from parameter and runtime-object reuse. The savings are larger on MoE, making Loop-PP 7.2\% and 0.9\% faster at PP2 and PP4. 
Phase profiling further shows that Loop-PP reduces all-to-all event duration from 4.195\,s to 3.451\,s 
at PP2 and from 4.498\,s to 3.779\,s at PP4.
This association may reflect runtime effects of parameter reuse, such as lower memory pressure and more favorable communication-buffer behavior.
Loop-PP therefore benefits MoE workloads more, where larger local savings better amortize the recurrent pipeline overhead.

\paragraph{Impact of Recurrent Structure}
\label{sec:recurrent-structure-results}

\begin{figure*}[t]
    \centering
    \includegraphics[width=\textwidth]{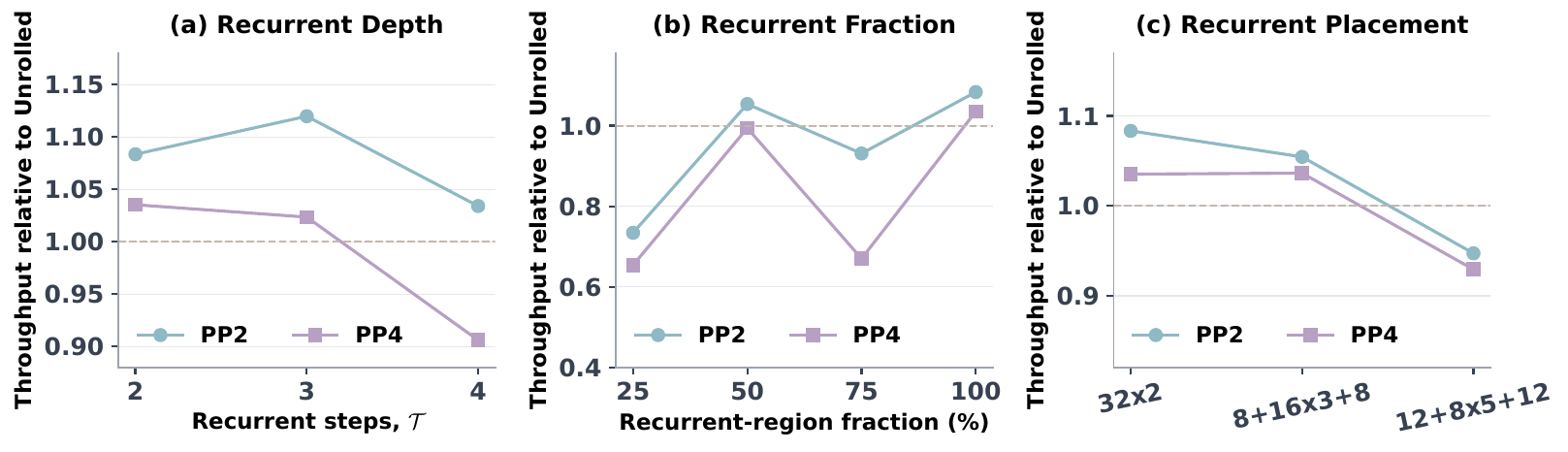}
    \caption{
    Sensitivity to recurrent structure on the 8B workload.
    (a) Throughput relative to Unrolled as full-loop recurrent depth increases.
    (b) Throughput relative to Unrolled across recurrent-region fractions at $\mathcal{T}=2$.
    (c) Three prelude--recurrent--coda layouts with the same 64-layer effective depth.
    }
    \label{fig:recurrent-structure-results}
\end{figure*}

We study recurrent structure on the 8B model by varying the number of recurrent
steps, the recurrent fraction, and the recurrent layout.
Figure~\ref{fig:recurrent-structure-results}(a) compares PP2 and PP4 as the
number of full-loop steps increases from two to four.
As $\mathcal{T}$ grows to four, Loop-PP's relative throughput weakens at both PP
degrees and falls behind Unrolled at PP4, showing that more recurrent
traversals incur a larger structural tax.
Figure~\ref{fig:recurrent-structure-results}(b) fixes $\mathcal{T}=2$
and varies the recurrent fraction from 25\% to 100\% at PP2 and PP4.
Performance varies non-monotonically with the recurrent fraction: the 50\% and
100\% configurations outperform the 25\% and 75\% configurations.
This shows that performance depends on how the recurrent region maps to
pipeline stages, rather than on recurrent fraction alone.
Figure~\ref{fig:recurrent-structure-results}(c) compares three layouts at the
same 64-layer effective depth.
The full-loop $32\times2$ and $8+16\times3+8$ layouts are close at PP4,
while $32\times2$ leads at PP2.
The $12+8\times5+12$ layout incurs more recurrent traversals and cross-stage
feedback, reducing opportunities to hide communication and
increasing dependency stalls.
Overall, more recurrent traversals and placement-sensitive cross-stage
dependencies amplify the structural tax, favoring pipeline layouts with
sufficiently thick compute chunks.

%% file: content/RW.tex
\section{Related Work}
\label{sec:related-work}

\paragraph{Pipeline Parallelism}
Conventional PP partitions layers across devices for feed-forward execution.
GPipe~\citep{huang2019gpipe} pipelines micro-batches, while PipeDream-2BW~\citep{narayanan2021pipedream2bw} 
and Megatron-LM~\citep{narayanan2021efficient} improve utilization through 1F1B scheduling, weight buffering, and interleaved virtual stages.
Subsequent work reduces pipeline bubbles through finer-grained or asynchronous scheduling~\citep{qi2024zerobubble,ajanthan2025nesterov,jung2026basisrotation,zmushko2026onestep,chen2026amdp}, 
improves memory efficiency~\citep{kim2023bpipe,qi2024controllable,wan2025pipeoffload}, or automates partitioning and scheduling~\citep{jhoo2025pfeife}.
GraphPipe~\citep{jeon2025graphpipe} and PULSE~\citep{sun2026pulse} extend PP to DAG-structured networks and skip-aware diffusion U-Nets, respectively, 
but do not directly address repeated execution of shared physical chunks with cross-stage recurrent feedback.
Other work pipelines computation along token, sequence, or time dimensions~\citep{li2021terapipe,zhao2026stateflow,bao2026tpipe}, 
which is orthogonal to our depth-recurrent setting.

\paragraph{Looped Transformers}
Looped Transformers increase effective depth by repeatedly applying shared layers. 
Building on early weight-tied architectures~\citep{dehghani2019universal,lan2020albert}, 
recent models reuse intermediate blocks or entire layer stacks~\citep{geiping2025huginn,zhu2025ouro,saunshi2025reasoning,kapl2026growing}, 
demonstrating benefits for reasoning, in-context learning, and test-time scaling~\citep{geiping2025huginn,zhu2025ouro,saunshi2025reasoning,yang2024looped,alabdulmohsin2025rins}.
Subsequent work explores relaxed weight sharing~\citep{bae2025relaxed}, 
adaptive recursion depth~\citep{bae2025mor,fu2025thinkathard,chen2025innerthinking,song2026adaponder,elbayad2020depthadaptive,jeddi2026loopformer,goyal2026elt}, 
dynamic skipping or looping~\citep{li2026polar}, 
and parallel test-time execution~\citep{wu2025parallelloop,yang2026loopcoderv2}.
Other studies investigate training stability~\citep{prairie2026parcae,fu2026simplystabilizing,movahedi2026fixedpoint,wang2026residualscaling}, 
cross-loop memory and attention costs~\citep{vendrell2026melt,oneill2026lla,deng2026lt2}, 
MoE-compatible recurrence~\citep{csordas2024moeut,lee2026sparselayers,chen2026loopmoe,gao2026loopies,knupp2026dreamer,chen2026mue}, 
and compute-matched scaling~\citep{schwethelm2026isodepth,prairie2026parcae,wang2026smelt}.
Despite these advances, pipeline-parallel training under cross-stage depth-recurrent feedback remains underexplored.

%% file: content/appendix.tex
\section{Semantic Graph Construction Details}
\label{app:semantic-graph}

\subsection{Formal Node and Edge Representation}

For micro-batch $m$, let $\mathcal{G}_m=(\mathcal{V}_m,\mathcal{E}_m)$ denote its finitely unrolled forward graph.
The Semantic Graph for $M$ micro-batches is their disjoint union,
\begin{equation}
    \mathcal{G}
    =
    \bigsqcup_{m=0}^{M-1}\mathcal{G}_m.
    \label{eq:semantic-graph-union}
\end{equation}
There are no model-semantic edges between different micro-batches; their interleaving is determined by the global compute scheduler.

A semantic node is represented as
\begin{equation}
    v =
    \bigl(
        m_v,\,
        r_v,\,
        t_v,\,
        j_v,\,
        p_v,\,
        q_v,\,
        [\ell_v^{\mathrm{start}},\ell_v^{\mathrm{end}})
    \bigr),
    \label{eq:semantic-node-record}
\end{equation}
where $m_v$ is the micro-batch, $r_v$ is the computation role, $t_v$ is the recurrent-step index, $j_v$ is the segment index, $(p_v,q_v)$ is the PP/VPP placement, and the final interval identifies the physical layer range when the node executes Transformer layers.
The recurrent-step index is undefined for non-recurrent nodes, and the layer interval is empty for control-only nodes.

Let $\mu(v)$ denote the physical module invoked by node $v$.
Logical visits at different recurrent steps are distinct nodes but may satisfy $\mu(v)=\mu(v')$.
They therefore execute with the same parameter object and accumulate gradients into the same physical parameter storage.

A semantic edge is represented as
\begin{equation}
    e =
    \left(
        u_e,\,
        v_e,\,
        \chi_e,\,
        \Sigma_e,\,
        \mathcal{B}_e
    \right),
    \label{eq:semantic-edge-record}
\end{equation}
where $u_e$ and $v_e$ are the producer and consumer, $\chi_e$ is the logical channel, $\Sigma_e$ is an ordered payload schema, and $\mathcal{B}_e$ maps each producer output slot to the corresponding consumer input slot.
This record refines the body shorthand $e=(u,v,\Sigma_e)$; below, $u$ and $v$ abbreviate $u_e$ and $v_e$ when the channel and slot binding are not germane.
Payload order is part of the communication contract.
The principal channels are \textsc{Main}, \textsc{Feedback}, and \textsc{Injection}; representative payloads include the hidden activation, loop state, residual source, core input, injection source, and model-specific precomputed state.
The schema associated with a backward transfer contains gradients for the same ordered payload fields.

\subsection{Derived Backward Semantics}

The Semantic Graph contains only forward model dependencies.
Native-Guided Schedule Synthesis derives a forward operation $(F,v)$ and a backward operation $(B,v)$ for every $v\in\mathcal{V}$.
For each forward edge $(u,v)\in\mathcal{E}$, it adds the forward dependency $(F,u)\rightarrow(F,v)$ and the reverse-mode dependency $(B,v)\rightarrow(B,u)$, together with $(F,v)\rightarrow(B,v)$.

For an output slot $k$ of $u$, let
$\operatorname{Out}(u,k)$ contain exactly the outgoing edges whose binding
$\mathcal{B}_e$ consumes that slot. Its backward operation consumes the sum of
the corresponding pullback contributions:
\begin{equation}
    \nabla o_{u,k}
    =
    \sum_{e\in\operatorname{Out}(u,k)}
    \nabla_e o_{u,k}.
    \label{eq:semantic-gradient-fanin}
\end{equation}
The runtime tracks this expected fan-in per producer slot and executes $(B,u)$ only after all required contributions are available.

\subsection{Semantic Equivalence}

\paragraph{Theorem.}
Assume that each logical visit executes the same deterministic kernel given the
same inputs, physical parameters, saved visit-local state, and logical
random-number stream as its counterpart in a statically tied
$\mathcal{T}$-step unrolling. During one training step, physical parameters
remain fixed, and any other model-visible mutable state is either fixed or
represented by semantic dependencies with the same update order. Assume also
that every semantic edge preserves its ordered tensor bundle and slot binding
exactly. Then any execution that respects the forward semantic dependencies
and their derived reverse-mode dependencies produces the same model outputs
and the same mathematical gradients for every input and shared physical
parameter as that tied unrolling. Floating-point gradient values may differ
only through permissible reordering of mathematically identical sums.

\paragraph{Proof.}
Order the forward nodes topologically.
The initial input and prelude nodes agree with the tied unrolling by construction.
Suppose all predecessors of node $v$ produce the same output bundles as their counterparts in the tied unrolling.
The schemas and slot bindings on the incoming semantic edges deliver exactly those bundles to the same physical kernel $\mu(v)$.
The kernel therefore produces the same output, establishing forward equivalence by induction.
In particular, the final coda output is identical.

For backward, traverse the derived reverse dependencies.
The output gradient is identical because the model output is identical.
For any node $u$ and output slot $k$,
Equation~\ref{eq:semantic-gradient-fanin} accumulates exactly one contribution
for every forward use selected by $\mathcal{B}_e$. Once these contributions
match as a mathematical sum, applying the same local backward kernel to the
same visit-local saved state yields the same input and physical-parameter
gradient contributions. Reverse induction therefore establishes equality of
every mathematical edge gradient. Summing the contributions from all logical
visits gives the same mathematical gradient for each shared physical
parameter; an implementation may accumulate those contributions in a
different floating-point order.
Replacing a local semantic edge with a communication transfer does not change this argument because the transfer preserves the schema and payload values by assumption.
\hfill$\square$

\section{Correctness of Compiled Streams}
\label{app:per-rank-compilation}

\paragraph{Planned-transfer semantics.}
The interpreter processes each per-rank stream in order: a \textsc{Compute}
action completes before the interpreter advances, a \textsc{POST} advances
only after every operation in its actual runtime action unit has been
successfully submitted to a backend execution context without waiting for its
peer, and a \textsc{WAIT} advances only after completion. Backend ordering
dependencies that precede progress eligibility of a submitted operation are
included in the augmented progress graph defined below. For every planned
transfer, the sender POST binds the payload
produced by its source compute, and the source buffer remains valid until the
sender WAIT completes. A transfer can complete only after its endpoint
operations have been issued. The destination WAIT returns only after that
transfer has completed and its ordered payload is available to the consumer.
Transport preserves the payload schema and values.

For communication matching, let $d(\gamma)$ denote the backend matching domain
of transfer $\gamma$, including its communicator, directed peer, and any
transport-specific tag namespace. Operations in distinct domains cannot match.
No unplanned P2P operation shares a domain with the compiled transfers during
their execution. Within one domain, transfer identities are unique, and the
backend either matches that identity explicitly or, as in the evaluated
ordered-P2P path, matches sends and receives in issue order. The latter path
therefore requires both endpoints to issue the common epoch-induced subsequence
for that domain.

\paragraph{Proposition 1 (Semantic-order preservation).}
Suppose compilation succeeds and the generated streams
$\{\mathcal{Q}_s\}_{s=0}^{P-1}$ pass the validations performed by the compiler after linearization. Then any completed
joint execution under the planned-transfer semantics respects every forward
and reverse-mode dependency of the Semantic Graph, as well as
forward-before-backward order for every semantic node.

\paragraph{Proof.}
Let $\prec_{\mathrm{exec}}$ denote happens-before in the completed joint
execution. Action coverage gives exactly one forward and one backward compute
action for every semantic node and the required endpoint actions for every
remote transfer. Because every $\mathcal{Q}_s$ is validated as a linear
extension of the Action DAG induced on stage $s$, each stage-local DAG edge is
preserved by stream order. The interpreter semantics then turns such stream
order into execution order: an earlier blocking action completes, or an earlier
POST is issued, before the interpreter starts the later action.

Consider an arbitrary semantic edge $e=(u,v,\Sigma_e)$. If $e$ is local, its
endpoints occupy the same stage. The Action DAG contains
\[
\operatorname{Compute}^{F}(u)
\prec
\operatorname{Compute}^{F}(v)
\quad\text{and}\quad
\operatorname{Compute}^{B}(v)
\prec
\operatorname{Compute}^{B}(u).
\]
Both are stage-local edges and hence hold under
$\prec_{\mathrm{exec}}$ by the validated linear-extension property.

If $e$ is remote, fix a phase $\phi$ and its planned transfer $\gamma$.
Write $(x,y)=(u,v)$ for $\phi=F$ and $(x,y)=(v,u)$ for $\phi=B$.
Equation~\ref{eq:remote-edge-actions} places the source compute before the
sender POST on the source stage and the destination WAIT before the consumer
compute on the destination stage. Stage-local validation and ordered
interpretation therefore give
\[
\operatorname{Compute}^{\phi}(x)\text{ completes}
\prec_{\mathrm{exec}}
\operatorname{POST}_{\mathrm{src}}(\gamma)\text{ is issued}
\]
and
\[
\operatorname{WAIT}_{\mathrm{dst}}(\gamma)\text{ returns}
\prec_{\mathrm{exec}}
\operatorname{Compute}^{\phi}(y)\text{ starts}.
\]
By planned-transfer causality, the transfer cannot complete before its sender
POST is issued, and its destination WAIT cannot return before completion.
Combining these relations yields
\[
\operatorname{Compute}^{\phi}(x)\text{ completes}
\prec_{\mathrm{exec}}
\operatorname{POST}_{\mathrm{src}}(\gamma)\text{ is issued}
\prec_{\mathrm{exec}}
\operatorname{transfer}(\gamma)\text{ completes}
\prec_{\mathrm{exec}}
\operatorname{WAIT}_{\mathrm{dst}}(\gamma)\text{ returns}
\prec_{\mathrm{exec}}
\operatorname{Compute}^{\phi}(y)\text{ starts}.
\]
Thus the forward dependency is preserved when $\phi=F$, and the derived
reverse-mode dependency is preserved when $\phi=B$.

Finally, for every semantic node $v$, the Action DAG contains the stage-local
edge
$\operatorname{Compute}^{F}(v)\prec\operatorname{Compute}^{B}(v)$,
which is preserved by the same argument. The local and remote cases exhaust
all semantic edges, so every claimed dependency holds.
\hfill$\square$

\paragraph{Proposition 2 (Communication matching safety).}
Suppose compilation succeeds, the generated streams pass the action-coverage,
FIFO, and mirrored-endpoint validations performed by the compiler after linearization, and each backend matching domain
obeys the matching semantics stated above. Then every send has exactly one
intended receive with the same transfer identity and payload schema, and no
send can cross-match with a different planned transfer.

\paragraph{Proof.}
Every remote semantic edge and phase creates one executable transfer record
$\gamma^+$ with identity determined by the unique pair $(e,\phi)$. Its identity
and payload schema are shared by the sender and receiver endpoint records.
Mirrored-endpoint validation requires one sender POST and one destination POST
for that same record, while complete action coverage rejects a missing
endpoint, a duplicate endpoint, or an endpoint that does not belong to a
planned transfer.

Fix an arbitrary backend matching domain $d$. Filter the global epoch order to
the transfers in that domain and denote the resulting sequence by
\[
L_d=[\gamma_0,\gamma_1,\ldots,\gamma_{q-1}].
\]
The epoch compiler inserts both endpoint POSTs of every $\gamma_i$ at the same
global epoch position. The per-rank projection advances its epoch cursor
monotonically. It may interleave local compute and WAIT actions, but cannot
skip or permute the epoch subsequence owned by either endpoint. The final FIFO
and mirrored-endpoint validations consequently establish that the sender
issues
\[
[\operatorname{send}(\gamma_0),\ldots,
  \operatorname{send}(\gamma_{q-1})]
\]
and the receiver issues
\[
[\operatorname{recv}(\gamma_0),\ldots,
  \operatorname{recv}(\gamma_{q-1})]
\]
in exactly the same order, with neither omissions nor duplicates.

If the backend uses explicit transfer identities in $d$, uniqueness and equal
endpoint identities force $\operatorname{send}(\gamma_i)$ to match
$\operatorname{recv}(\gamma_i)$. Otherwise the backend uses FIFO matching.
For $i=0$, the first unmatched operations at the two endpoints are the send
and receive of $\gamma_0$, so they match. Assume inductively that the first
$i$ pairs have matched as intended. The first unmatched operations are then
the send and receive of $\gamma_i$; FIFO matching pairs them, and they cannot
match any $\gamma_j$ with $j\neq i$. Induction proves intended matching for
all $q$ transfers in $d$. Since distinct matching domains cannot cross-match,
applying the argument independently to every $d$ proves the global
one-send/one-receive and no-cross-match properties. Finally, the paired
endpoints are projections of the same $\gamma^+$, so their transfer identity
and ordered payload schema agree. The logical-channel FIFO validation further
preserves the channel-specific order required by the consumer.
\hfill$\square$

\paragraph{Augmented progress graph.}
Here an action is an actual runtime action unit after communication packing,
native-table projection, and any static repair; operations submitted atomically
by one packed POST share one POST milestone. For every such action $a$ in a
final stream, let $\widehat{a}$ denote the milestone
at which its rank may advance: completion for \textsc{Compute} and control
actions, successful issue for \textsc{POST}, and return for \textsc{WAIT}.
The augmented progress graph $\mathcal{G}_{\mathrm{prog}}$ contains one such
milestone for every action. For each consecutive pair
$a_{s,i},a_{s,i+1}$ in $\mathcal{Q}_s$, it contains the local stream edge
\[
    \widehat{a}_{s,i}\longrightarrow\widehat{a}_{s,i+1}.
\]
For every planned transfer $\gamma$, it additionally contains
\[
\begin{split}
 \widehat{\operatorname{POST}}_{\mathrm{src}}(\gamma),
 \widehat{\operatorname{POST}}_{\mathrm{dst}}(\gamma)
 &\longrightarrow
 \widehat{\operatorname{WAIT}}_{\mathrm{src}}(\gamma),\\
 \widehat{\operatorname{POST}}_{\mathrm{src}}(\gamma),
 \widehat{\operatorname{POST}}_{\mathrm{dst}}(\gamma)
 &\longrightarrow
 \widehat{\operatorname{WAIT}}_{\mathrm{dst}}(\gamma).
\end{split}
\]
These edges model the fact that either endpoint completion may require both
endpoint operations to have been issued. Importantly, the local edges are
derived from the \emph{final} runtime streams, after all transformations that
can coalesce or reorder operations. The compiler checks
$\mathcal{G}_{\mathrm{prog}}$ for acyclicity in
$O(|V|+|E|)$ time before execution.

\paragraph{Proposition 3 (Conditional PP-transport deadlock freedom).}
Suppose the generated streams pass the validations performed by the compiler after linearization and
$\mathcal{G}_{\mathrm{prog}}$ is acyclic. Assume a fair execution in which
(i) every enabled compute or control action terminates in finite time,
(ii) a \textsc{POST} submits its packed operations and returns in finite time
without waiting for its peer, (iii) once both endpoint POSTs of a planned
transfer have become progress-eligible,
the transport completes that transfer in finite time even if an endpoint
interpreter is waiting, and (iv) the compiled PP runtime introduces no
additional blocking synchronization outside $\mathcal{G}_{\mathrm{prog}}$.
Then the compiled PP action streams contain no circular wait and complete in
finite time; this claim does not extend to failures or synchronization outside
the compiled PP runtime.

\paragraph{Proof.}
Assume for contradiction that a fair execution reaches a permanent state with
at least one unfinished milestone. The subgraph of
$\mathcal{G}_{\mathrm{prog}}$ induced by the unfinished milestones is a
nonempty finite DAG, so it has a vertex $\widehat{a}$ with no unfinished
predecessor. Hence every predecessor of $\widehat{a}$ has completed.
Its local-stream predecessor, if any, has therefore completed, so $a$ is the
next action of its rank.

If $a$ is a compute or control action, fairness and assumption~(i) imply that
it completes. If $a$ is a POST, fairness and assumption~(ii) imply that it is
issued. If $a$ is a WAIT, both endpoint POST milestones are predecessors of
$\widehat{a}$ and have completed, including their modeled backend-ordering
predecessors; assumptions~(iii) and fairness therefore imply that the transfer
completes and the WAIT returns. In every case an
unfinished milestone advances, contradicting permanence of the assumed state.
Thus no permanent blocked state with unfinished actions is reachable.
Because the streams are finite, repeated progress completes all of them.
\hfill$\square$

\paragraph{Consequence and scope.}
Proposition~1 establishes semantic dependency order, and Proposition~2
discharges the transport-matching premise for the actual endpoint operations.
Together with the assumptions of the Semantic Equivalence theorem in
Appendix~\ref{app:semantic-graph}, they establish forward and backward semantic
preservation for every completed joint execution. Proposition~3 separately
establishes PP-transport deadlock freedom under the stated fairness and
transport-progress assumptions. The compiler validates the structural
premise---acyclicity of the augmented progress graph over the final runtime
action units---but cannot validate backend progress, hardware failure freedom,
or synchronization introduced outside the compiled PP runtime.